\documentclass{aa501}
\usepackage{graphicx}
\begin{document}

\title{Axisymmetrical Gas Inflow in the Central Region of NGC 7331}

\author{E. Battaner \inst{1}, E. Mediavilla \inst{2}, A. Guijarro
          \inst{3}, S. Arribas \inst{4} \and E. Florido \inst{1}}

\offprints{E. Battaner, battaner@ugr.es}

\institute{Dpto. F\'{\i}sica Te\'orica y del Cosmos, Universidad de Granada,
          Spain
          \and Instituto de Astrof\'{\i}sica de Canarias, Tenerife,
          Spain
          \and Centro Astron\'omico Hispano Alem\'an, Almer\'{\i}a, Spain
          \and Space Telescope Science Institute, Baltimore, USA. Affiliated with the RSS Department of the European Space Agency. On leave from the IAC - CSIC}		
\date{}
\authorrunning{Battaner et al.}
\titlerunning{Axisymmetrical Gas Inflow in the Central Region of NGC 7331}

\abstract{New Integral Field Spectroscopy
of the central region of NGC 7331 reveals strong H$\alpha$ emission in
the well-known CO and HI ring of NGC 7331. The [NII]/H$\alpha$ ratio
indicates that a large scale stellar formation process is taken place
at the ring in agreement with previous hypothesis about the exhaustion
of gas in the inner to the ring region. The dynamics of stars and gas
are not coupled. There is a ring of peculiar velocities in the ionized
gas velocity map. These peculiar velocities can be well interpreted by
the presence of an axisymmetric inflow of 40 km/s at the inner
boundary of the large-scale gaseous ring. We infer an inwards total
flux of 1.6 M$_{\odot}$yr$^{-1}$. This value is typical of the
accretion rates in hypothetical {\bf large} nuclear black holes. Despite the large
differences in the scales of the nucleus and the gas ring of NGC 7331,
we suggest that this inwards flux is feeding the nucleus.  
}

\maketitle

\section{Introduction}
Peculiar motions are frequently detected in active galaxies where
the existence of several kinematically distinct gaseous systems, some of
them suffering radial outward movements, has been attributed to the
influence of the active nucleus (Garc\'{\i}a-Lorenzo et al. 1999,
2001, Arribas et al. 1997, Mediavilla \& Arribas 1993). The outflow of gas
close to the nucleus is relatively easy to detect because of the
large velocities involved (of the order of 100 km/s) and of the
direct illumination from the strong active source. Inflow of gas seems
to be much more difficult to observe as it is presumably caused by
less outstanding mechanisms than the nuclear activity. However, it is
very important to detect since inflow should provide the matter that
will eventually accrete into the nuclear black hole. To study the
influence of the environment on the nucleus we will present new
results about the kinematics in the central region of NGC 7331. Some
peculiarities previously detected in the ionized gas velocity field of
this galaxy (see below) make it specially interesting for this kind of
studies.

The large scale gas distribution in NGC 7331 is ring-like as first detected by
Bosma (1978) in HI, Telesco et al. (1982) from NIR colors and Young and
Scoville (1982) in CO. The ring is confirmed in other wavelengths and is taken
today as a prototype gas ring.

Bower et al. (1993) used optical long-slit
spectra to show that their models without a central black hole fit
the observational data. However, there is an unresolved LINER nucleus
(see, for instance, Cowan et al., 1994), so this galaxy could harbour a
massive black hole of $5\times10^8 M_{\odot}$. This is supported by
the motion of ionized gas ([NII] + H$\alpha$ emission line) reported by
Afanasiev et al. (1989) in the 0.2-0.4 kpc (2.8-5.6 arcsec) zone and
by the discovery of a nuclear X-ray source by Stockdale et al. (1998)
using ROSAT. 

The existence of a bar has been suggested from
non-circular motions by Marcelin et al. (1994) and von Linden et
al. (1996). However the I and K band photometry in the central regions
carried out by Prada et al. (1996) did not indicate any significant
barred morphology.

In the central kpc, Bottema (1999)
found that the emission line gas (for R $\leq$ 40 arcsec) seems to
rotate slower than the stars. An explanation of the observations
would consist in an inclined and warped gaseous plane.  

Using 2D spectroscopy, Mediavilla et al. (1997) concluded that the
kinematics of the stars and most of the ionized gas are
decoupled. These authors found that the kinematic axes of the ionized
gas velocity map are distorted and rotated with respect to the stars,
something that can be interpreted in terms of radial
movements. However, the reduced field of view of the available velocity maps
(about 7'' $\times$  7'') makes difficult to verify this interpretation. 

In this letter we are going to present new Integral Field spectroscopy
covering the central 30'' $\times$ 30'' of NGC 7331, with the aim of
understanding the kinematics of the region surrounding the nucleus.   

\section{Observations and reduction}

NGC 7331 was observed with INTEGRAL (Arribas et al. 1998) in combination with
the fiber spectrograph WYFFOS (Bingham et al. 1994) the 15 of July of 2001. The
spectral resolution and coverage were 4.8\AA\ and 1445\AA\, respectively. The
central wavelength was 6200\AA . From the wavelength calibration we estimate
that the uncertainty in velocity determinations is better than 10 km/s. We used the
SB3 bundle of INTEGRAL. This bundle consists of a central rectangle of 30''
$\times$ 34'' and an outer ring of 90'' in diameter. Four exposures of 600s
each were taken from the central region of NGC 7331. These frames were combined
to increase the S/N ratio and to reject the cosmic rays. Posteriorly, the
spectra were extracted, calibrated in wavelength, and corrected from
throughput. The reduction steps were made using the INTEGRAL data reduction
package running in IRAF. More details about the reduction procedure can be
found in Arribas et al. (1991).

For several spectral features of interest we have determined a grid of
values at the locations of the fibers at the focal plane and have
interpolated a map of the selected feature (see, e.g., Arribas et al. 1999). 

\begin{figure*}
   \includegraphics{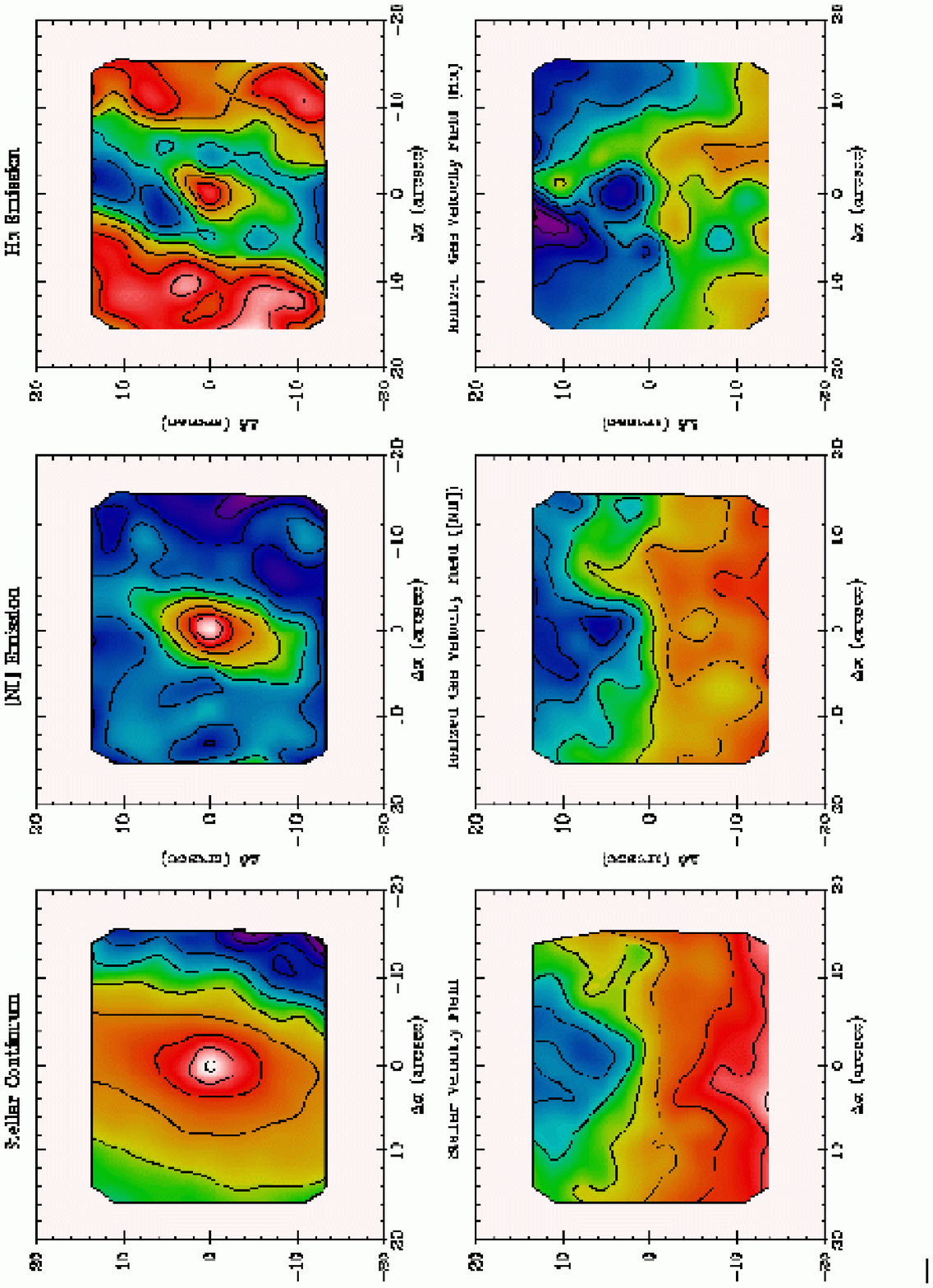}
   \caption{Intensity and velocity maps for NGC 7331 (see text). The intensity maps isophotes (arbitrary units) are logaritmically scaled. The steps between consecutive isophotes are 0.31, 0.18, and 0.25 for the continuum, [NII] and H$\alpha$ maps, respectively. The step between two consecutive iso-velocity lines is 35 km/s. Approaching (receding) velocities correspond to blue (red) colors. The dashed isovelocity line corresponds to $v=$830 km/s, the systemic velocity according to Mediavilla et al. (1997)}
   \label{fig:maps}
\end{figure*}

In Fig. \ref{fig:maps} we show the resulting intensity and velocity maps. The
stellar continuum corresponds to the $\sim$ 5730-6800\AA, spectral range. The
stellar velocity map has been obtained by cross correlating the spectra in the
$\sim$ 5600\AA-6850\AA\ wavelength interval using the nuclear spectrum as
template.  We also present intensity and velocity maps obtained from the
multi-Gaussian fit of the H$\alpha$+[NII]$\lambda\lambda 6548,6583$ blend. The
nuclear spectrum is very similar to the ones obtained from the same region by
Filippenko and Sargent (1985) or Mediavilla et al. (1997) and shows a
strong H$\alpha$ absorption (Mediavilla et al. give a description of
the nuclear spectrum of NGC 7331 in comparison with that of M31). To
avoid the influence of the strong H$\alpha$ absorption in the determination of the
velocity and intensity maps, we have considered in the fits an extra component
representing this absorption feature (see Mediavilla et al. 1997, for details).

\section{Observational results}

A very noticeable feature in Fig. \ref{fig:maps} is the ring
structure observed in the H$\alpha$ intensity map. The two-dimensional
distribution of the H$\alpha$ emission consists of a nucleus, a region
with very low emission and then the H$\alpha$ ring that begins rather
abruptly at about 7 arcsec measured in the east direction ($\sim$ 500 pc
equivalent to a galactocentric radius of about 2 kpc, considering $i =
75^o$). Therefore we are observing the inner part of the well-known
large-scale gaseous ring. In contrast, the [NII] emission decays
continuously from the nucleus towards the outer parts and is very
faint at the ring location. In the ring the [NII]$\lambda
6583$/H$\alpha$ ratio is about 0.3, typical of ionization by a
HII-like continuum. This indicates that the ring is representative of
a large scale stellar formation process. On the contrary, the
[NII]$\lambda 6583$/H$\alpha$ ratio in the inner to the ring region is very
high, about 5. This is typical of LINERS. At the nucleus we found a value
[NII]$\lambda 6583$/H$\alpha\sim 3$ also typical of LINERS and similar to the
one (2.7) reported by Mediavilla et al. (1997). It is also remarkable
that the transition between the LINER and the starburst regions is very
abrupt and takes place at the inner boundary of the H$\alpha$ ring.

Another important feature in the maps of Fig. \ref{fig:maps} is
the presence of a symmetrical distortion in the velocity maps
corresponding to the ionized gas, also coincident with the inner
boundary of the starburst ring.  
 
Leaving aside the S-shaped distortion present in the innermost region
(see Mediavilla et al. 1997), the stellar velocity map in NGC 7331 is
quite regular. However, the dynamics observed
in H$\alpha$ and [NII] is very much distorted when compared with the pure
rotation velocity field of the stellar system. Notice that the central 7''
$\times$ 7'' of the H$\alpha$ map (Fig. \ref{fig:maps}) are in very
good agreement with the H$\alpha$ velocity map presented in Mediavilla et
al. (1997). 

\section{The large scale inflow}

\begin{figure}
   \centering
   \includegraphics[width=8cm]{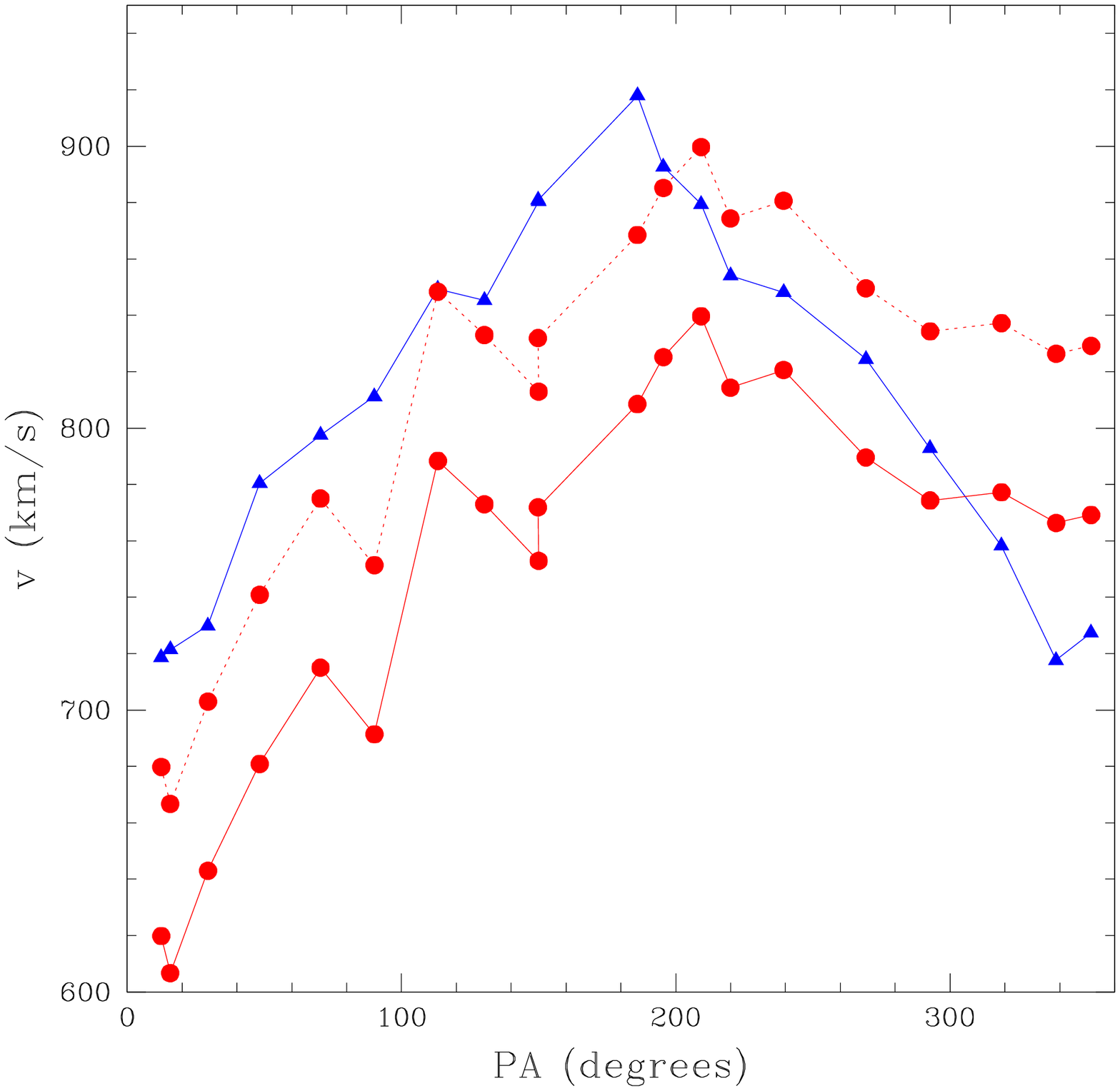}
   \caption{$v(\theta)$ along the peculiar velocities ring. Triangles
   correspond to the stellar component. Circles joined by a solid line
   correspond to H$\alpha$. Circles joined by a dot line correspond to
   H$\alpha$ corrected from blueshift (see text).}
   \label{fig:velo}
\end{figure}

The new map of H$\alpha$ velocities (Fig. \ref{fig:maps}) gives a global
perspective of the central region of NGC 7331 and can be interpreted
in terms of a ring of peculiar velocities superimposed to a normal
rotation pattern.  In Fig. \ref{fig:velo}
we represent the velocity as a function of galactic azimuthal angle,  $v(\theta)$, along the ring of peculiar
velocities for H$\alpha$ and the stars. As it can be seen the H$\alpha$
curve presents a global blueshift of 60 km/s with respect to the stellar curve
(the [NII] curve, not represented here, suffers also a blueshift of 23 km/s).
After removing the global blueshift, the H$\alpha$ curve is shifted by
about 25 degrees with respect to the stellar one (see Fig.
\ref{fig:velo}). The [NII] curve corrected from blueshift is
intermediate between that of the stars and H$\alpha$ showing a shift
of 15 degrees with respect to the stellar rotation. The amplitudes of
the H$\alpha$ and [NII] curves are 117 and 95 km/s,
respectively. Thus, the kinematic shifts with respect to the stellar
rotation imply a radial velocity component of about 55 km/s for
H$\alpha$ and 26 km/s for [NII].  With an alternative method,
measuring the departures of the velocity field from the systemic
velocity along the minor kinematic axis, we have obtained a similar
value, 45 km/s for H$\alpha$. 

\begin{figure}
   \centering
   \includegraphics[width=5cm]{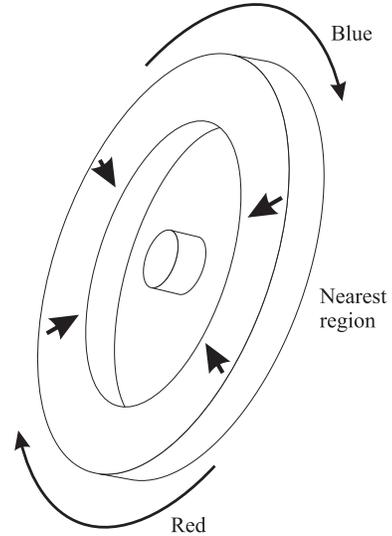}
   \caption{A scheme of the nearly axisymmetric gas flow from the
   H$\alpha$ ring to the central inner region detected in
   NGC7331. West is to the right and North is up. The internal radius
   of the plotted ring is about 2 kpc.}
   \label{fig:ring}
\end{figure}

The west part of this kinematically peculiar ring is relatively redder
and the east-part relatively bluer. As the dust lanes in optical images
clearly indicate that the west part is the closest, we can conclude that
we are observing a contraction motion (see Fig.
\ref{fig:ring}). Thus, we have detected gas flowing from the H$\alpha$
ring to the central inner region with a very noticeable
axisymmetry. The velocity is maximum at the inner boundary of the
large-scale gaseous ring. It then decreases towards the center,
probably as V $\propto$ (R $\rho$)$^{-1}$, where $V$ is the inward
component of the gas velocity at the inner boundary of the ring, $R$
the galactocentric radius and $\rho$ the density, due to continuity requirements.

{\bf Interpretations other than axisymmetric inflow cannot be in principle 
disregarded to explain the radial motion. An elliptical motion would produce a
departure from circular motion, without a net inflow. Also, the observed
motions could be attributable to a pair of tightly wound spiral arms, i.e.
streaming motions along the spiral arms due to a density wave could produce a
similar observed velocity pattern. These effects are not axisymmetric, but as
long as axisymmetry is not perfect, they should be considered as alternative
interpretations. However, it would be more difficult to account for the
observed global blueshift in the framework of these alternative explanations.
In any event, we favor axisymmetric inflow as the simplest explanation.}

Regarding the global blueshift, it is noticeable that the magnitudes of the
vertical (responsible of the global blueshift) and radial inward (causing the
kinematic axis shifts) motions are in good quantitative agreement in both
cases, H$\alpha$ and [NII]. This suggests that we are seeing gas flowing from
the edge of the H$\alpha$ starburst ring in all allowed directions. Thus, the
global blueshift is explained by the same process that gives rise to the radial
inward motion. 

There is good spatial coincidence between the peculiar velocities and H$\alpha$
starburst rings (particularly good at the east region where the edge of the
H$\alpha$ ring is more sharply defined). Anyway, there is not significant
differences in the results about the global blueshift and the radial inward
motion at the galaxy plane if the velocity curves are determined along the
H$\alpha$ ring. The intermediate position of the [NII] kinematics between the
stars and H$\alpha$ can be explained if more than one kinematic component is
contributing to the [NII] emission. This is supported by the existence of
two-peaked [NII] lines (Mediavilla et al. 1997).

The inward velocity is much greater than the sound speed and a pressure
gradient can be ruled out as the mechanism causing the gas
inflow. However, it is of the order of the Alfven velocity if we
consider a magnetic field of $10^{-5}$ Gauss. That means that the
inflow of gas from the ring to the center could be induced by magnetic
field pressure forces if a strong but in principle acceptable magnetic
field were associated to the ring (see Battaner et
al. 1988). Alternatively, the inflow can be also attributed to the
presence of strong winds originated in starbursts. The existence of
a large scale process of stellar formation in NGC 7331 was suggested
by Young \& Scoville (1982) to explain the presence of the CO ring in
terms of exhaustion of the gas to form stars in the central
region. The discovery of large scale stellar formation in the ring
fits naturally in this hypothesis that is also supported by studies of
the Mg2 index (del Burgo et al. 1997).

The gas inflow from the ring could eventually be accreted into the central
black hole. To do an estimation of the accreted mass rate, we consider
that the inwards flux is 2$\pi$RHV$\rho$. The gas density ($\rho$) is taken as
0.5$\times$10$^{-2}$ M$_{\odot}$pc$^{-3}$ (Israel \& Baas 1999), the
equivalent gas disc height as H = 467 pc (Bottema 1999) and the
galactocentric radius as R = 2 kpc. Therefore we estimate the infalling flux as 1.6
M$_{\odot}$yr$^{-1}$ of the order of magnitude of the accretion mass
rate in a  luminous seyfert nuclear black hole (Blandford et
al. 1990). 

For a low luminosity AGN, such as that in M81, the required accretion
flux has been estimated to be of the order of
$5 \times 10^{-5}$ M$_\odot$/yr (Ho et al. 1997), assuming an
efficiency of conversion between matter
and energy of 0.1. As NGC 7331 has a nucleus of 
comparable luminosity (Stockdale et al. 1998) it seems reasonable that
a part of the radial macroscopic infall observed here can feed the
NGC7331 AGN.

Considering the large difference between the scales of the nuclear
accretion disc and the macroscopic ring observed here, the
interpretation of the inward flux as the feeding of the central black
hole can only be considered as tentative. Actually, this identification
should be assessed under the conditions of continuity and stationarity.
Even the inward
velocity, as a function of radius, peaks at the inner ring boundary,
it is not vanishing at lower radii. This fact favours the
interpretation that the observed flux eventually reach the nucleus.

\section{Conclusions}

Some conclusions are worthy to be summarized:

We have detected an H$\alpha$ ring coincident with the inner part
of the well known CO and HI ring. The [NII]/H$\alpha$ ratio at the
ring is typical of HII-like ionization and indicates that a large
scale stellar formation process is taken place at the ring. This is in
agreement with the hypothesis of that the gas in the inner to the ring
region has been exhausted by massive stellar formation, that continues
at the ring. The physical conditions in the inner to the ring region
are typical of LINERS and the transition between starburst and LINER
takes place very sharply at the inner boundary of the H$\alpha$ ring.  

Ionized gas peculiar velocities are found near to the inner boundary of
the CO-HI ring, compatible with an axisymmetric morphology
superimposed to the otherwise regular map of the ionized gas. The
comparison with the regular stellar velocity map makes the assumption
of an axisymmetric contraction very plausible. The infalling matter is
observed at the inner boundary of the large-scale gaseous ring and
likely continues towards the inner region. Although several mechanisms
are possible to explain the origin of the inflow, the existence of
strong winds at the ring wall fits naturally with the presence of
massive stellar formation in the ring.

Even if the accreting flux could have very different
causes we would be observing infalling velocities of the order of 50
km/s and infalling gaseous fluxes of the order of 1
M$_{\odot}$yr$^{-1}$.  This is higher than the
required accretion rate for a central black hole. Probably
we are observing the infalling motion feeding the black hole
having its source as far as 2 kpc away from the center.

The inwards flux axisymmetry seems to rule out, in this galaxy, 
mechanisms based on bars. Bars have been proposed as sources
for feeding AGN's and this have been observed to operate in the central
regions of some active galaxies (See Perez et al. 2000, and references
therein). NGC 7331, without a clearly observed bar, and with an
axisymmetric inflow suggest that other ways of loosing angular
momentum could also be important.

\begin{acknowledgements}

The Isaac Newton Group of Telescopes (ING) operates the 4.2m William
Herschel Telescope on behalf of the Particle Physics and Astronomy 
Research Council (PPARC) of the United Kingdom and the Netherlands
Organization for Scientific Research (NWO) of the Netherlands. The ING
is located at the Roque de Los Muchachos Observatory, La Palma, Spain. 

This paper has been in part supported by the ``Plan Andaluz de Investigaci\'on''
(FQM-108) and by the ``Secretar\'{\i}a de Estado de Pol\'{\i}tica
Cient\'{\i}fica y Tecnol\'ogica'' (AYA2000-1574).

 We acknowledge the improvement of the paper by an anonymous referee.

\end{acknowledgements}

\end{document}